\documentclass{elsart5p}
\usepackage{epsfig}
\usepackage{epsfig}
\usepackage{bm}
\usepackage{color}

\usepackage[latin1]{inputenc}
\newcommand{\beq}{\begin{equation}}
\newcommand{\eeq}{\end{equation}}
\newcommand{\bea}{\begin{eqnarray}}
\newcommand{\eea}{\end{eqnarray}}

\usepackage{amsmath}
\usepackage{amsfonts}
\usepackage{amssymb}
\usepackage{bm}
\usepackage{color}
\usepackage{graphicx}

\newcommand{\ee}{\end{equation}}
\newcommand{\be}{\begin{equation}}

\begin{document}

\begin{frontmatter}

\title{Fisher equation with turbulence in one dimension.}


\author{Roberto Benzi}
\address{Dip. di Fisica, Univ. di Roma "Tor Vergata", via della Ricerca Scientifica 1, 00133, Roma, Italy}
\author{David R. Nelson}
\address{Lyman Laboratory of Physics, Harvard University, Cambridge, Ma 02138 U.S.A.}

\begin{abstract}
We investigate the dynamics of the Fisher equation for the spreading of micro-organisms in one dimenison  subject to both turbulent convection and diffusion. 
We show that for strong enough
turbulence, bacteria , for example, track in a  quasilocalized fashion (with remakably long persistance times) sinks in the turbulent field. 
An important consequence is a large reduction in the carrying capacity of the fluid medium. We determine analytically the
regimes where this quasi-localized behavior occurs and test our predictions by numerical simulations.

\end{abstract}
\end{frontmatter}

The spreading of bacterial colonies at very low Reynolds numbers on
 a Petri dish can often be described \cite{Wakika} by the Fisher equation
\cite{fisher}, i.e.
\be
\label{fisher_0}
\partial_t c  = D \partial_{xx}^2 c + \mu c - bc^2 ,
\ee
where $c(x,t)$ is a continuous variable describing the concentration of micro-organisms, 
$D$ is the diffusion coefficient and $\mu$ the growth rate.

In the last few years, a number of theoretical and
experimental studies \cite{shapiro}, \cite{mutsu}, \cite{benjacob}, \cite{nelson1}, \cite{nelson2} have been performed to understand the spreading and extinction of a population in 
an inhomogeneous environment.  In this paper we study  a particular time-dependent inhomogeneous enviroment, namely
the case of the field $c(x,t)$  subject to both convection and diffusion and satisfying the equation:
\be
\label{fischer}
\partial_t c + div( \vec{U} c) = D \nabla^2 c + \mu c - bc^2
\ee
where $U(x,t)$ is a turbulent velocity field. Upon specializing to one dimension, we have
\be
\label{1d}
\partial_t c + \partial_x (U c)  = D \partial_{x}^2 c + \mu c- b c^2
\ee
Equation (\ref{1d}) is relevant for the case of compressible flows, 
where $\partial_x U \ne 0$, and for the case
when the field $c(x,t)$ describes the population of 
inertial particles or biological species.
For  inertial particles,
it is known \cite{inertial}  that for large Stokes number , i.e. the ratio between the characteristic particle response time and 
the smallest time scale due to the hydrodynamic viscosity,
the flow advecting $c(x,t)$ is effectively  compressible, even if the 
particles move in an incompressible fluid. 
Let us remark that the case of compressible turbulence
is also 
relevant in many astrophysical applications where (\ref{fischer}) is used as a simplified
prototype of combustion dynamics.   By suitable rescaling of $c(x.t)$, we can always set $b=1$. In the following, unless stated otherwise, 
we shall assume $b=1$ whenever $\mu \ne 0$ and $b=0$ for $\mu=0$. For a treatment of equation (\ref{1d}) with a spatially uniform but
time-dependent random velocity, see \cite{xx}

\bigskip

\begin{figure}
\centering
\epsfig{width=.50\textwidth,file=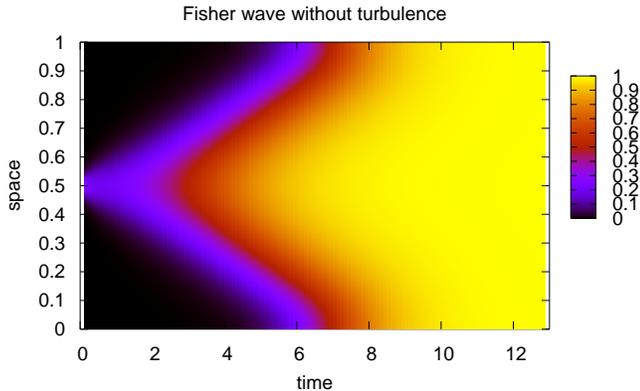}
\caption{  Numerical simulation of eq. \ref{fisher_0} with $\mu=1$, $D=0.005$ and with periodic boundary conditions. The initial conditions are $c(x,t)=0$ everywhere expect for few grid points near $L/2=0.5$ where $c=1$. The horizontal axis represents time while the vertical axis is space. The colors display different contour levels of $c(x,t)$.}
\label{fig01}
\end{figure}

The Fisher equation has travelling
front solutions that propagate with velocity $v_F \sim (D\mu)^{1/2}$ \cite{fisher}, \cite{kpp}. 
In Fig. (\ref{fig01}) we show a numerical solution of Eq.  (\ref{fisher_0}) with $D=0.005$, $\mu= 1$ obtained by numerical
integration on a space domain of size $L=1$ with periodic boundary conditions. The figure shows the space-time behaviour
of $c(x,t)$, the color code representing the curves $c(x,t) = const$. With initial condition $c(x,t=0)$ nonzero  on only a few grid points centered at $x=L/2$, $c(x,t)$
spreads with a velocity $v_F \sim 0.07 $ and, after a time $L/v_F \sim 4$ reaches the boundary.  

\begin{figure}
\centering
\epsfig{width=.50\textwidth,file=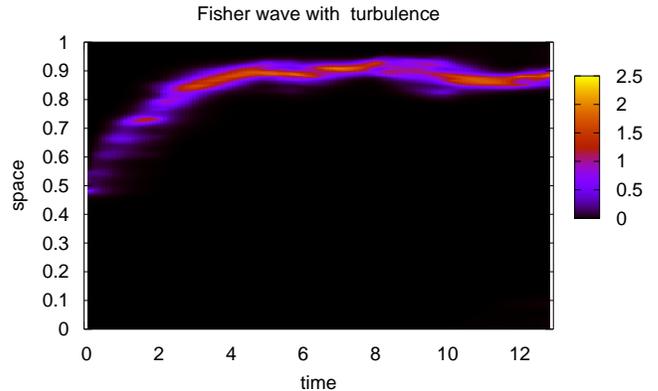}
\caption{  Same parameters and initial condition  as in Fig. (\ref{fig01}) for equation (\ref{1d}) with a "strong turbulent"  flow $u$ advecting  $c(x,t)$.}
\label{fig02}
\end{figure}

A striking result, which motivated  our investigation, is displayed in Fig. (\ref{fig02}), showing
the numerical solutions of  Eq. (\ref{1d}) for a relatively "strong" turbulent flow, where the average convection velocity vanishes and
"strong turbulence"  means high Reynolds number ( a more precise definition of the Reynolds number and specification of the velocity field is given
in the following sections). From the figure we see no trace of a propagating front: instead, a well-localized pattern of $c(x,t)$ forms and stays more or less in a stationary
position.

\begin{figure}
\centering
\epsfig{width=.50\textwidth,file=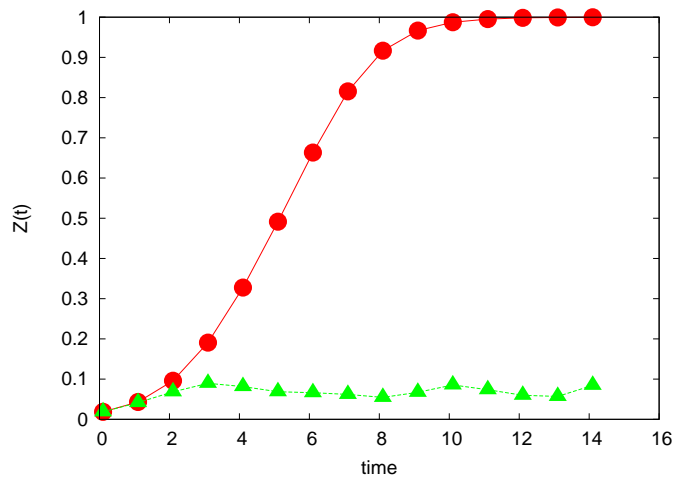}
\caption{  The behavior in time of the total "mass" $Z(t) \equiv \int dx c(x,t)$. The red circles show the function $Z$ for the case of Fig. (\ref{fig01}), i.e. a Fisher wave
with no turbulence. The green triangles show $Z$ for the case of Fig. (\ref{fig02}) when a strong turbulent flows is advecting $c(x,t)$. }
\label{fig03}
\end{figure}

For us, Fig. (\ref{fig02}) shows a counter intuitive result. One naive expectation might be  that 
turbulence enhances mixing. The mixing effect due to turbulence is usually parametrized in the literature \cite{frisch}
by assuming an effective (eddy)  diffusion coefficient $D_{eff} \gg D$.
As a consequence, one  naive guess for Eq. (\ref{1d}) is that the spreading of an initial population is qualitatively similar to the travelling
Fisher wave with a more diffuse interface of width $\sqrt{D_{eff}/\mu}$.  As we have seen,  this naive prediction is wrong for strong enough turbulence: the solution of
equation (\ref{1d}) shows  remarkable localized features which are preserved on time scales longer than the characteristic growth time $1/\mu$ or even the Fisher wave propagation time
 $L/v_F$. 
An important consequence of the localization effect is that the global "mass" (of growing microorganisms, say) ,  $Z \equiv \int dx c(x,t)$,  behaves differently with and without turbulence.
In Fig. (\ref{fig03}), we show $Z(t)$:
the curve with red circles refers to the conditions shown in Fig. (\ref{fig01})),
while the curve with green triangles to Fig. (\ref{fig02}).  

The behavior of $Z$ for the Fisher equation without turbulence is a familiar {\it S}-shaped curve that reaches
the maximum $Z=1$ on a time scale $L/v_F$. On the other hand,  the effect of turbulence (because of
localization)   on the Fishe equation dynamics reduces significantly $Z$ almost by one order of magnitude. 
  
With biological applications in mind, it is important to determine conditions such that the spatial distribution of microbial organisms and the carrying capacity of the medium
are significantly altered by convective turbulence.
Within the framework of the Fisher equation, 
localization effect has been studied for a constant convection velocity and quenched time-independent spatial dependence in the
growth rate $\mu$  \cite{nelson1}, \cite{nelson2},  \cite{nelson3}, \cite{neicu}. In our case, localization,
when it happens, is  a time-dependent feature and depends on the statistical properties of the compressible turbulent flows. 
As discussed in detail below, a better term for the phenomenon we study here might be "quasilocalization", in the sense that (1) spatial localization of the growing
population sometimes occurs at more than one location; (2) these spatial locations drift slowly about and (3) localization is intermittent in time, as localized populations collapse and 
then reform elsewhere.
For these reasons, the quasilocalization studied here is not quite the same phenomena as the Anderson localization of electrons in a disordered potential studied in \cite{anderson}.
Nevertheless, the similarities are sufficiently strong that we shall use the terms "quasilocalization" and "localization" interchangeably in this paper.
It is worth noting that the localized "boom and bust" population
cycles studied here may significantly effect 
 "gene surfing"  \cite{genesurf} at the edge of a growing population, i.e. by changing the probability of gene mutation and fixation in the population.

For the case of bacterial populations subject to both turbulence and convections due to, say,
an external force such as sedimentation under the action of gravity, we may think that the turbulent velocity
can be decomposed into a constant "wind" $u_0$ and a turbulent fluctuation $u(x,t)$ with zero mean value, $U(x,t) = u_0 + u(x,t)$.  We find that
the localization
shown in Fig. (\ref{fig02}) can be significantly changed for large enough background convection $u_0$.

We would like to understand why and how $u_0 \ne 0$ can change the statistical properties of $c(x,t)$ in the presence of a random convecting velocity field. We wish to understand,
in particular,  whether $c(x,t)$ spreads or localizes as a function of parameters such as
the  turbulence intensity and the  mean "wind" speed $u_0$. 


Our results are based on a number of numerical simulations of Eq. (\ref{1d}) performed
using a particular model for the fluctuating velocity field $u(x,t)$. In Sec. $1$ we introduce the model and we
describe some details of the numerical simulations. In Sec. $2$ we develop a simple
"phenomenological" theory of the physics of Eq.  (\ref{1d}) based on our present understanding of turbulent
dynamics. In Sec. $3$ we analyze the numerical results when the  sedimentation  velocity $u_0=0$
while in 
Sec. $4$ we describe our findings for $u_0 > 0$ . Conclusions follow in Sec. $5$.

\section{The model}

To completely specify equation (\ref{1d}) we must define the dynamics of the "turbulent" velocity field
$U(x,t)$. For now, we set $U(x,t) = u_0+u(x,t)$, neglect the uniform part $u_0=0$ and focus on $u(x,t)$.
Although we consider a one dimensional case, we want to study the
statistical properties of $c(x,t)$ subjected to turbulent fluctuations which are close
to thos generated by the three dimensional 
Navier-Stokes equations. Hence, the statistical properties of $u(x,t)$ should be
described characterized by intermittency both in space and in time.  We
build the turbulent field $u(x,t)$  by appealing to a simplified  shell model of fluid
turbulence \cite{biferale}. The
wavenumber space is divided into shells of scale $k_n = 2^{n-1} k_0$, $n=1,2,...$. For each shell with
characteristic wavenumber $k_n$,  we describe 
turbulence by using the complex Fourier-like variable $u_n(t)$, satisfing the following equation of motion:
\bea
  (\frac{d }{dt}&+&\nu k_n^2 ) u_n = i(k_{n+1} u_{n+1}^* u_{n+2}-\delta
  k_n u_{n-1}^* u_{n+1} \nonumber  \\
&+&(1-\delta) k_{n-1} u_{n-1} u_{n-2})+f_n \ . 
\label{sabra}
\eea
The model contains one free parameter, $\delta$,
and it conserves two quadratic invariants (when the force and the
dissipation terms are absent) for all values of $\delta$. The first is
the total energy $\sum_n |u_n|^2$ and the second is $\sum_n (-1)^n
k_n^{\alpha} |u_n|^2$, where $\alpha= \log_{2} (1-\delta)$.  
In this
note we fix $\delta = -0.4$. For this value of $\delta$ the model
reproduces intermittency features of the real three dimensional Navier Stokes equation with surprising
good accuracy \cite{biferale}. Using $u_n$, we can build the real one dimensional velocity field $u(x,t)$ as follows:
\be
u(x,t) = F\sum_n [ u_n e^{i k_n x} + u^*_n e^{-i k_n x} ],
\ee
where $F$ is a free parameter to tune the strength of velocity fluctuations
(given by $u_n$) relative to other parameters in the model (see next section).
In all numerical simulations we use a forcing function  $f_n = (\epsilon(1+i)/u^{*}_1 )\delta_{n,1}$, i.e. energy is supplied only
to the largest scale corresponding to $n=1$.
With
this choice, the input power in the shell model is simply given by $1/2\sum_n [u^{*}_n f_n + u_n f^{*}_n] = \epsilon$ , i.e. 
it is constant in time.  
To solve Eqs. (\ref{1d}) and (\ref{sabra})  we use a finite difference scheme with periodic boundary conditions.

Theses model equations can be studied in detail without
major computational efforts.  One main point of this note is to explore the qualitative
and quantitative dynamics of Eqs.  (\ref{1d}),(\ref{sabra}) and compare it against the
phenomenological theory developed in the next section.

The free parameters of the model are the diffusion constant $D$, the size of the periodic 1d spatial domain $L$,
the growth rate $\mu$, the viscosity $\nu$ (which fixes the Reynolds number
 $Re$), the mean constant velociy $u_0$,
the ``strength' of the turbulence $F$ and finally the power input in the shell model, namely 
$\epsilon$. Note that according to the Kolmogorov theory \cite{frisch}, $\epsilon \sim u_{rms}^3/L$ where $u_{rms}^2$ is 
the mean square velocity. Since $u_{rms} \sim F$, we obtain that $F$ and $\epsilon$ are related as $\epsilon \sim F^3$.
By rescaling of  space, we
can always put $L=1$.  We fix $\epsilon=0.04$ and $\nu= 10^{-6}$, corresponding to an equivalent
$Re = u_{rms} L/\nu  \sim 3  \times 10^{5}$.
As we shall see in the following, most of our numerical results are
independent of $Re$ when $Re$ is large enough. In the limit $Re \rightarrow \infty$,  the
statistical properties of eq. (\ref{1d}) depend on 
the remaining free parameters, $D$, $u_0$, $\mu$ and $F$. The important combinations of these parameters are discussed in the next section.

\section{ Theoretical considerations}

We start our analysis by rewriting (\ref{1d}) in the form:
\be
\label{1d_new}
\partial_t c + (u_0+w)\partial_x (c)  = D \partial_x^2 c + (\mu+g)c- bc^2
\ee
where  $w\equiv u(x,t)$ and $g(x,t) \equiv -\partial_x u(x,t)$. Previous theoretical
investigations \cite{nelson1} have shown that for $u_0 = w = 0$, $c(x,t)$ becomes localized in space for
time-independent "random" forcing $g=g(x)$ (Anderson localization \cite{anderson}). 
For $u_0$ large enough, a transition from localized 
to extended solutions has been predicted and observed in previous numerical and theoretical
works \cite{nelson3}. Here, we wish  to understand whether something resembling localized solutions survives in equation (\ref{1d_new})
when both $w$ and $g$ depend on time as well as space. 

To motivate our subsequent analysis, consider first
the case $\mu=u_b=0$. In this limit, Eq. (\ref{1d}) is just the Fokker-Planck
equation describing the probability distribution $P(x,t) \equiv c(x,t) $ to find a particle in the range $(x, x+dx)$ at time $t$, whose
dynamics is given by the stochastic differential equation:
\be
\label{particle}
\frac{dx}{dt} = u(x,t) + \sqrt{ 2D} \eta(t) 
\ee
where $\eta(t)$ is a white noise with $\langle \eta(t)\eta(t')\rangle = \delta(t-t')$. Let us assume for the moment that $u(x,t)=u(x)$ is time independent. Then,
the stationary solution of (\ref{1d}) is given by 
\be
\label{7}
P(x,t) = A^{-1} exp[-\Phi(x)/D]
\ee
where $A$ is a normalization constant and
$\partial_x \Phi = -u(x)$. It follows that $P(x,t)=P(x)$ is strongly peaked near the points $x_i$
where
$\Phi$ has a local minimum, i.e. $u(x_i) = 0$ and $-\partial^2_{xx} \Phi \equiv \partial_x u(x)|_{x=x_i} < 0$. 
Let us now consider the behaviour of $P(x)$ near one particular point $x_0$ where $u(x_0)= 0$. For
$x$ close to $x_0$ we can write:
\be
\label{x0}
\frac{dx}{dt} = - \Gamma_0 (x-x_0) + \sqrt{2D} \eta(t)
\ee
where $\Gamma_0 \equiv -\partial_x u(x)|_{x=x_0}$.
Equation (\ref{x0}) is the Langevin equation for an overdamped harmonic oscillator, and 
tells us that $P$ is spread around $x_0$ with a characteristic "localizaiont length" 
of order $\xi_l \equiv \sqrt{D/\Gamma_0}$. On the other hand, we can identify  $\Gamma_0$ with $\Gamma$, a typical  gradient of the 
turbulent velocity field $u$. In a turbulent flow, the velocity field is correlated over spatial scale
of order $v_*/\Gamma$ where $v_*^2/2$ is the average kinetic energy of the flow. For $P$ to
be localized near $x_0$, despite spatial variation in the turbulent field,  we must require that the localization length $\xi_l$ should be smaller
than the turbulent correlation scale $v_*/\Gamma$, i.e.
\be
\sqrt{\frac{D}{\Gamma}} < \frac{v_*}{\Gamma} \rightarrow \frac{v_*^2}{D\Gamma} > 2
\label{prima}
\ee
Condition (\ref{prima}) can be easily understood by considering the simple case of a periodic velocity field
$u$, i.e. $u = v_* cos(x v_* /\Gamma)$. In this case, condition (\ref{prima}) states that
$D$ should be small enough for the probability $P$ not to spread over all the minima of $u$.
For small $D$ or equivalently for large $v_*^2/\Gamma$, 
the solution will be localized near the minima
of $u$, at least for the case of a frozen turbulent velocity field $u(x)$.

\bigskip

The above analysis can be extended for velocity field $u(x,t)$ that depend on both space and time. 
The crucial observation is
that,  close to the minima $x_i$ of $\Phi(x,t) \equiv - \int dx u(x,t)$, we should have 
$u(x_i,t) \sim 0$. 
Thus, 
although $u$ is a time dependent function, sharp peaks in 
 $P(x,t)$ move quite slowly, simply because $u(x,t) \sim 0$ near the maximum of $P(x,t)$. 
 One can consider
a Lagrangian path $x(t)$ such that $x(0)=x_0$, where $x_0$ is one particular point 
where $u(x_0,0)=0$ and $\partial_x u(x,0)|_{x=x_0} < 0$.
From direct numerical simulation of Lagrangian particles in fully
developed turbulence, we know that the acceleration of Lagrangian particles is a  strongly 
intermittent quantitiy, i.e. it is small most of the time with large (intermittent) bursts. Thus,
we expect that the localized solution of $P$ follows $x(t)$ for quite long times except for
intermittent bursts in the turbulent flow. During such bursts, the position where
$u=0$ changes abruptly, i.e. almost discontinuosly from one point, say $x(t)$, to
another point $x(t+\delta t)$. During the short time interval $\delta t$, $P$ will drift and spread,
eventually reforming to become  localized again near $x(t+\delta t)$.
The above discussion suggests that the probability $P(x,t)$ will be
localized most of the  time in the Lagrangian frame, except for
short time intervals $\delta t$ during an intermittent burst.

\bigskip

We now revisit the condition (\ref{prima}). For the case of a time-dependent
velocity field $u$, we estimate $\Gamma$ as the characteristic gradient of the velocity field,
i.e. 
$$
\Gamma \sim \langle (\partial_x u)^2 \rangle^{1/2}
$$
where $\langle .. \rangle$ stands for a time average. Now, $v_*^2$ should be considered as
the mean kinetic energy of the turbulent fluctuations.
In our model, both $v_*$ and $\Gamma$ are
proportional to $F$, the strength of the velocity fluctuations. Thus, we can rewrite the localization criteria (\ref{prima}) in the form:
\be
\label{11}
\frac{v_*^2F}{D\Gamma} > 2
\ee
where $v_*$ and $\Gamma$ are computed for $F=1$. We conclude that for small 
values of $F$, $P(x,t)$ is spread out,  while for large $F$, $P$ should be a localized or sharply peaked function
of $x$ most of the time. An abrupt transition, or at least a 
sharp crossover, from extended to sharply peaked functions  $P$, should
be observed for increasingf $F$.

\bigskip 

It is relatively simple to extend the above analysis for a non zero growth rate $\mu > 0$. 
The requirement (\ref{prima}) is now only
a necessary condition to observe localization in $c$. For $\mu>0$ we must also require
that the characteristic gradient on scale $\xi_l$ must be larger than $\mu$, i.e. the effect
of turbulence should act on a time scale smaller than $1/\mu$. We  estimate the gradient on scale
$\xi_l$ as $\delta v(\xi_l)/\xi_l$, where $\delta v(\xi_l)$ is
the characteristic velocity difference on scale $\xi_l$. We invoke 
the Kolmogorov theory, and set
$\delta v(\xi_l) = v_* (\xi_l/L)^{1/3}$ to obtain:
\be
\mu < \frac{\delta v(\xi_l)}{\xi_l} = \frac{v_* \xi_l^{-2/3}}{L^{1/3}} = v_*(\frac{\Gamma}{LD})^{1/3}
\label{seconda}
\ee
In (\ref{seconda}), we interpret $\Gamma $ as the characteristic velocity gradient of the turbulent
flow.  Because $v_* \sim F$ and $\Gamma \sim F$, it follows that the r.h.s of (\ref{seconda})
goes as $F^{4/3}$. Note also that $\delta v(\xi_l)/\xi_l \le \Gamma$ 
on the average, which leads to the inequality:
\be
\mu < \Gamma
\label{due}
\ee
From (\ref{prima}) and
(\ref{due}) we also find
\be
\frac{v_*^2}{D \mu} > 2
\ee
a second necessary condition. Once again, we see that  localization in a Lagrangian frame
should be expected for strong enough turbulence.

One may wonder whether a non zero growth rate  $\mu$ can change our previous conclusions about
the temporal
behavior, and in particular about its effect on the dynamics of the Lagrangian points where
$u(x,t)=0$. Consider the solution of (\ref{1d}) at time $t$, allow for a spatial domain of size $L$, and introduce
the average position
\be
\label{xm}
x_m \equiv \int_0^L dx x \frac{c(x,t)}{Z(t)}
\ee
where $Z(t)=\int_0^L dx c(x,t)$.
Upon assuming for simplicity a single localized solution,
 we can think of $x_m$ just as the position where most of the bacterial concentration $c(x,t)$ is localized. Using 
Eq. (\ref{1d}), we can compute the time derivative $v_m(t) = dx_m/dt$. After a short computation, we obtain:
\be
\label{vm}
v_m(t) = Z \int_0^L dx (x_m-x) P(x,t)^2 + \int_0^L u(x,t) P(x,t) dx
\ee
where $P(x,t) \equiv c(x,t)/Z(t)$  and $Z(t) = \int_0^L c(x,t) dx$. Note that $v_m$ is
independent of $\mu$. Moreover,
 when $c$ is localized near
$x_m$, {\it both} terms on the r.h.s. of (\ref{vm}) are close to zero. Thus, 
$v_m$ can be significantly different from zero only if
$c$ is no longer localized and the first integral on the r.h.s becomes relevant.
We can now understand
the effect
of the non linear term in (\ref{1d}): when $c(x,t)$ is localized, the non linear term is almost irrelevant
simply because $v_m$ is close to $0$. On the other hand, when
$c(x,t)$ is extended the non linear term drives the system to  the state $c=1$ which is an exact solution in the 
absence of turbulent convection
 $u(x,t)=0$. 

\bigskip

We now allow a non zero mean flow $u_0 \ne 0$. 
As  before, we first set
 $\mu = b=0 $ and consider a time -independent velocity field $u(x)$. Since the solution $c(x,t)$ of
(\ref{1d}) can still be interpreted as be the probability to find a particle in the interval $[x,x+dx]$ at time $t$,
we can rewrite (\ref{x0}) for the case $u_0 > 0$ as follows:
\be 
\frac{dx}{dt} = -\Gamma_0(x-x_0) + u_0 + \sqrt{2D} \eta(t)
\label{x0u0}
\ee
The solution of (\ref{1d}) is localized near the point 
$x_1=x_0+u_0/\Gamma_0$. Thus for small $u_0$ or large $\Gamma_0$ there is no major change in the
arguments leading to (\ref{prima}). In general, we expect that $P(x)$ will be localized near $x=x_0$,  provided 
the length $\xi_0 \equiv |x_1-x_0| = u_0/\Gamma_0 $ is smaller than $\xi_l$, i.e.
\be
\label{terza}
\frac{u_0}{\Gamma_0} < \sqrt{\frac{D}{2\Gamma_0}} \rightarrow \frac{u_0^2}{D\Gamma_0} < 2
\ee
When (\ref{terza}) is satisfied, then our previous analysis on localized solutions for both
$\mu=0$ and  $\mu \ne 0$ is still valid. Let us note that by combing (\ref{due}) and
(\ref{terza}) we obtain
\be
\label{nelson}
\frac{u_0^2}{D \mu} < 2
\ee
as a condition for localization, obtained in the study of localized/extended transition for steady flowsin an Eulerian context \cite{nelson2}, \cite{nelson3}.
Here we remark that in a turbulent flow, Eq. (\ref{nelson}) is only a necessary condition, because
(\ref{prima}),(\ref{seconda}) and (\ref{terza}) must all also be satisfied for $c$ to show quasi-localized
states. 

To study the change in the spatial behaviour of $P$ as a function of time,  we need a measure of the degree of localization.
to look for an some kind of order parameter.
Although there may be a number of valuable solutions,  an efficient measure should be related to the "order"/"disorder"
features of $c(x,t)$, where "order" means quasi-localized and "disorder" extended. 
As pointed out in the introduction, the total "mass" of the organisms $Z(t) \equiv \int_0^L dx c(x,t)$ is strongly affected by a strongly
peaked (or quasi-localized) $c(x,t)$, as opposed to a more extended concentration field.
However, a more illuminating quantity, easily studied in simulations, is 
\be
\label{entropia_s}
S(t) = - \int_0^L dx P(x,t) log(P(x,t))
\ee
where $P(x,t) \equiv c(x,t)/Z$.
Localized solutions of Eq. (\ref{1d}) correspond to small values of this entropy-like quantity  while extended solutions correspond
to large values of $S$, which can be interpreted as the information contained in the probability distribution $P(x,t)$ at time $t$.
In our numerical simulations, we consider a discretized form of 
(\ref{entropia_s}), namely:
\be
\label{entropia}
S(t) = - \sum_{i=1,N} \frac{c(x_i,t)}{Z} log(\frac{c(x_i,t)}{Z})
\ee
where $x_i$ are now the $N$ grid points used to discretize (\ref{1d}), $c(x_i,t)$ is the
solution of (\ref{1d}) in $x_i$ at time $t$ and $Z(t) \equiv \sum_i c(x_i,t) $.

\begin{figure}
\centering
\epsfig{width=.50\textwidth,file=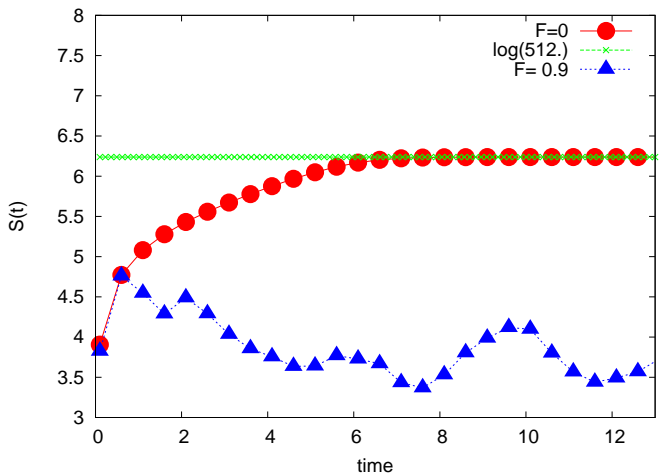}
\caption{  The behavior in time of  $S(t) $ . The red circles show the function $S(t)$ for the case of Fig. (\ref{fig01}), i.e. a Fisher wave
with no turbulence. The blue triangles show $S(t)$ for the case of Fig. (\ref{fig02}) when a "strong" turbulent flows is advecting $c(x,t)$. }
\label{fig1aa}
\end{figure}

To understand how well $S(t)$  describes whether $c(x,t)$ is localized or extended, 
we consider the cases discussed in the Introduction in Fig.s (\ref{fig01}) and (\ref{fig02}).
The numerical computations were done with $F=0$ for Fig. (\ref{fig01}) and $F=0.9$ for Fig. (\ref{fig02}), i.e. no turbulence
and "strong" turbulence (the attribute "strong" refers to the conditions (\ref{prima}) and (\ref{seconda}). In Fig. (\ref{fig1aa}) we show
$S(t)$ corresponding to the two simulations, namely $F=0$ (red circles) and $F=0.9$ (blue triangles). The initial condition is the same
for both simulations $c(x,0) = exp[-(x-L/2)^2/0.05]$, i.e. a rather localized starting point. It is quite clear, from inspecting Fig. (\ref{fig1aa}), that
$S(t)$ is a rather good indicator to detect whether $c(x,t)$ remains localized or becomes extended. While for $F=0$ (a quiescent fluid), $S$ reaches its maximum value( $S=9log(2.)$ 
for $512$ grid points)
at $t=6.$ (corresponding to uniform concentration $c(x,t)=1$), for $F=0.9$, $S$ is always close to its initial value $S \sim 4.$, indicating that $c(x,t)$ is localized, in agreement
with Fig. (\ref{fig02}).

\bigskip
Let us summarize our findings:  when subjected to turbulence, we  expect $c(x,t) $ to be "localized", i.e. strongly peaked, most of the time for large enough
$F$ and $u_0=\mu=b=0$. Upon  increasing the growth rate $\mu$, the value of $F$ where $c(x,t)$ shows Lagrangian
localization should 
increase. Finally, for fixed $F$ and $\mu$ we should find a localized/extended crossover for
large enough values of $u_0$. Because our theoretical analysis is based on scaling arguments, we
are not able to fix the critical values for which localized/extended transition should occour as
a function of $D$,$F$ and $u_0$. However, we expect that the 
conditions (\ref{prima}),(\ref{seconda}) and (\ref{terza})
capture the scaling properties in the parameter space of the model. Finally, we
have introduced an entropy like quantity $S(t)$ useful for analyzing
the time dependence of
of $c(x,t)$ and for distinguishing between localized and extended solutions. In the following section,
we compare our theoretical analysis against numerical simulations.

\bigskip

\section{ Numerical results for  $u_0=0$}

We now  discuss numerical results obtained by integrating equation Eq. (\ref{1d}). As discussed
in Sec. $1$, all numerical simulations have been done using periodic boundary conditions.
Eq. (\ref{1d}) has been discretized on a regular grid of $N=512$ points. Changing the resolution $N$, shifting $N$ to $N=1024$
or $N=128$, 
does not change the results discussed in the following. We use the same extended  initial condition $c(x,t)=1$ for
all numerical simulations with the few exceptions which discussed in the introduction (the Fisher wave) and in the
conclusions. For all simulations studied here,
the diffussion constant $D$ has been kept fixed at $D=0.005$.

\bigskip

\begin{figure}
\centering
\epsfig{width=.50\textwidth,file=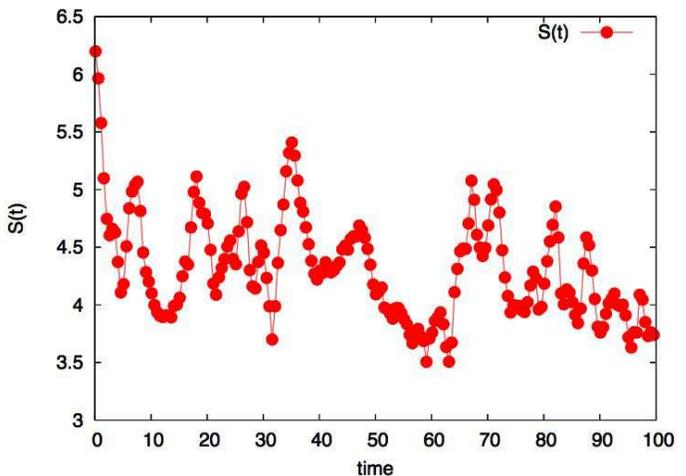}
\caption{  The behavior of $S(t)$ for a numerical simulation of (\ref{1d}) for $\mu= b=u_0=0$, no saturation term $c^2(x.t)$ and $F=0.5$. With our uniform
initial condition, $S(0) = 9log(2.)$. 
As discussed in Sec. 2, 
 $S(t)$ is a reasonable indicator for localized/extended spatial
behavior of $c(x,t)$. Since the flow is turbulent, $S(t)$ behaves chaotically.  However, it fluctuates at values
lower that $S(t=0)$ and indicates the degree of localization.
}
\label{fig1a}
\end{figure}
\begin{figure}
\centering
\epsfig{width=.50\textwidth,file=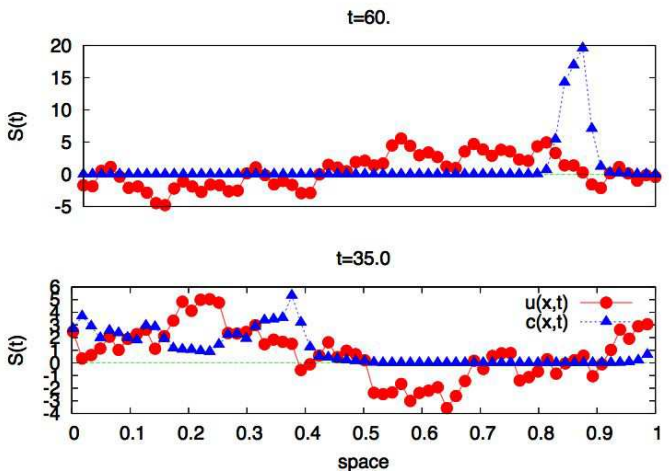}
\caption {  Numerical simulation  of Eq. (\ref{1d}) for $\mu= u_0=b=0$, term and $F=0.5.$. The upper panel shows $c(x,t)$ (blue triangles) and $u(x,t)$ (red circles) multiply by $10$, at
$t=60.$. The lower panel shows the same quantities at $t=35.$. The two time frames have been chosen to illustrate localized (upper panel) and more extended (lower panel) solutions.
Note that the localized solution at $t=60.$ reaches its maximum value for $u=0$ and $\partial_x u|_{u=0} <0$, as predicted by the analysis of sec. $2$.
}
\label{fig1b}
\end{figure}

We first discuss  the case $\mu= b=0$  term and beginby understanding how well $S(t)$ describes
the localized/extended feature of the $c(x,t)$.
In Fig. (\ref{fig1a}) we plot $S(t)$ as a function
of time for a case with $F=0.5.$. The behaviour of $S(t)$ is quite chaotic, as expected.  In Fig. (\ref{fig1b}) we
show the functions $c(x,t)$ for two particular times, namely $t=35.$ (lower panel)  and $t=60$ (upper panel). These two particular configurations
correspond to extended ($t=35$) and quasi-localized ($t=60$) solutions. The corresponding values of $S$ are $S=5.5$
for $t=35.$ and $S=3.5$ for $t=60$.
It is quite clear that for small $S$ strong localization characterizes $c(x)$ while for increasing 
$S$ the behaviour of $c(x)$ is more extended.

In Fig. (\ref{fig1b}) we also show (red circles) the instantaneous behavior of $u(x,t)$ (multiply by a factor $10$ to make
the figure readable). As one can see, the maximum of $c(x,t)$ always corresponds to points where $u=0$.

\begin{figure}
\centering
\epsfig{width=.50\textwidth,file=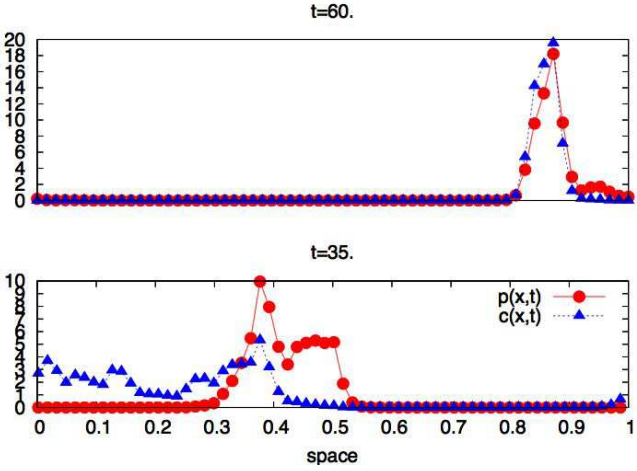}
\caption {  Numerical simulation  of (\ref{1d}) for $\mu=b = u_0=0$ and $F=0.5.$. The upper panel shows $c(x.t)$ (red circles) at $t=60.$and the behavior of $P(x,t)$ (blue triangle)
computed using (\ref{7}) using the instantaneous velocity field shown in Fig. (\ref{fig1b}). In the lower panel we show the same quantities ($c(x,t)$ and $P(x,t)$) at time $f=35.$
when the solution is extended. 
}
\label{fig1c}
\end{figure}

To understand whether the analysis
of Sec. 2 captures the main features of the dynamics. we plot in FIg. (\ref{fig1c}), for $t=35$ (lower panel) and $t=60$ (upper panel),
the quantity $P(x,t)$ as computed from Eq. (\ref{7}), i.e. by using
the instantaneous  velocity field $u(x,t)$. Although
there is a rather poor agreement between $c(x,t)$ and $P(x,t)$ at $t=35$, at time $t=60$ the $P(x,t)$ is a rather
good approximation of $c(x,t)$, i.e. when $c(x,t)$ is localized.  The spatial behavior of $c(x,t)$ is dictated by the point $x_0$ where $u=0$ and the velocity gradient
$\partial u|_{x=x_0}$ is large and negative.   All the above results are in qualitative agreement with our analysis.

\begin{figure}
\centering
\epsfig{width=.50\textwidth,file=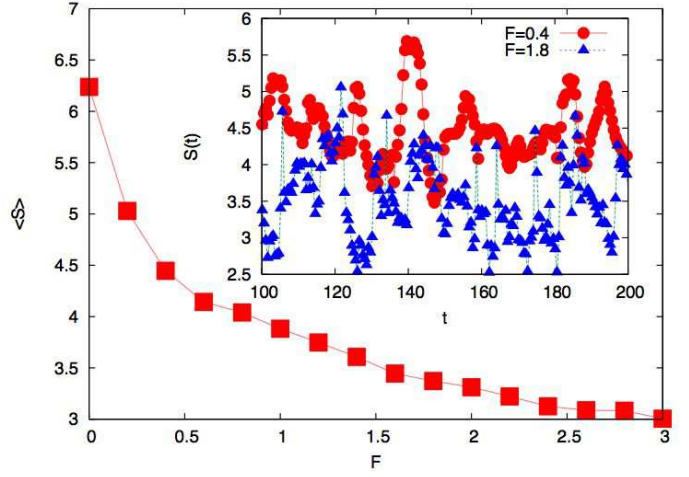}
\caption{ Time averaged entropy  $\langle S \rangle$ as a function of $F$. For large $F$ the system fluctuates about
small values of $S$,i.e.
$c(x,t)$ becomes more localized. In the insert, we show the time behaviour of $S(t)$ for two particular values of $F$,
namely $F=0.4$ (red curve) and $F=1.8$ (blue curve). Numerical simulations performed for $\mu=b=u_0=0$}
\label{fig2}
\end{figure}

Next we test the condition (\ref{11}), which states that localization should become more pronounced
for increasing values of $F$. To test Eq. (\ref{11}) we performed a number of numerical
simulations with long enough time integration to reach statistical stationarity. In Fig. (\ref{fig2})
we show $ \langle S \rangle$ as function of $F$, where $\langle ... \rangle$ means a time average.
In the insert of the same figure, we show to time dependence of $S(t)$ for two different values of $F$,
namely $F=0.4$ and $F=1.8$. The behavior of $\langle S \rangle$ is decreasing as a function of $F$,  in
agreement with (\ref{11}). The temporal behavior of $S$, for two individual realizations shown in the insert, reveals that, while on
the average $S$ decreases for increasing $F$, there are quite large oscillations in $S$, i.e. the system
shows both localized and extended states during its time evolution. However,
for large $F$ localization is more pronounced
and frequent. On the other hand, for small values of $F$, localization is
a "rare" event.  Overall, the qualitative picture emerging from Fig. (\ref{fig2}) is in agreement
with Eqs. (\ref{prima}) and (\ref{11})

\begin{figure}
\centering
\epsfig{width=.50\textwidth,file=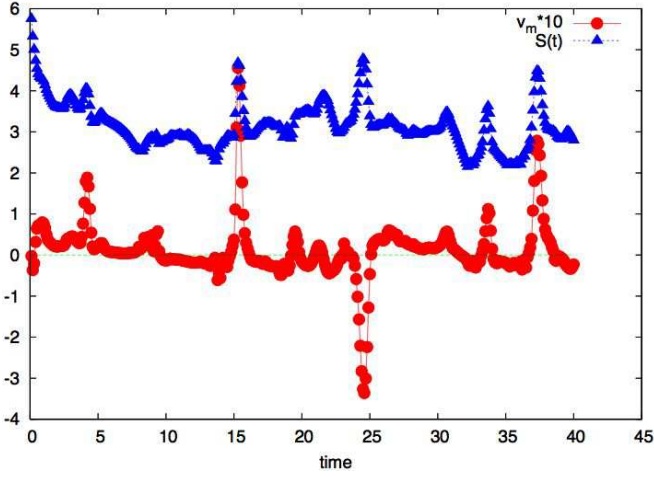}
\caption{ Time dependence of $v_m$ (red curve) and $S(t)$ (blue curve) for $F=2.0$. The value of $v_m$ is
multiplied by $10$ to make the figure readable. The velocity of the accumulation point for the bacterial concentration $c(x,t)$, 
$v_m$ , is computed using (\ref{vm}). We again set $\mu=u_0=b=0$.}
\label{fig3}
\end{figure}

In the previous section we argued that (\ref{prima}) and (\ref{11}) apply also
for a time-dependent function velocity $u(x,t)$. 
The basic idea was that $c(x,t)$ is localized near some point $x_0$ which
slowly changes in time, except for intermittent bursts. In sufficiently large systems, localization
about multiple points is possible as well.
During the intermittent burst, $c(x,t)$ spreads 
and after the burst $c(x,t)$ becomes localized around a new position $x_0$. 
We have already shown, in Figs.
(\ref{fig1b}) and (\ref{fig1c}), that our argument seems to be in agreement with the numerical computations using a time-dependent
velocity field.
To better understand this point,
 we measure  $v_m$ defined in Eq.
(\ref{vm}). We expect a small
 $v_m$  during localized epochs
 when $S$ is small. 
Each
time interval when $c$ is localized, should end and start with an intermittent burst where
$|v_m|$ may become large. Figs (\ref{fig3}) illustrates the above dynamics.
The solid red curve is $v_m(t)$ multiplied by a factor $10$ while the blue dotted curve shows $S(t)$. The 
numerical simulation is for $F=2.$, i.e. to a case where localization is predominant in the
system. Fig. (\ref{fig3}) clearly shows the "intermittent" bursts in the velocity $v_m$. The stagnation point velocity, 
punctuated by large positive and negative excursions,
typically wanders near $0$. If we assume a single sharp maximum in $c(x,t)$, as in the 
upper panel of  Fig. (\ref{fig1b}), 
the localized profile $c(x,t)$ does not move or  moves
quite slowly. During an intermittent burst, $v_m$ grows significantly while $c(x,t)$ spreads over the space.
Soon after the intermittent burst (see for instance the snapshot at time $t=15$ in Fig. (\ref{fig3})),
the velocity $v_m$ becomes small again and the corresponding value of $S$ decreases. Fig.
(\ref{fig3}) provides a concise summary  of the dynamics:  both localized and extended configurations
of $c(x,t)$ are observed as a function of time. During a era of localization, a bacterial concentration described by $c(x,t)$
is in a kind 
of "quasi-frozen" configuration.

Fig. (\ref{fig3}) tells us that condition (\ref{prima}), which was derived initially for a frozen
turbulent field $u$,  works as well for time dependent turbulent fluctuations. As $F$ increases,
 the
system undergoes a sharp crossover and the dynamics of $c(x,t)$ slows down in localized configurations.
Additional features  of this  transition will be discussed later on when we focus on a quantity analogous to the specific heat.

\begin{figure}
\centering
\epsfig{width=.50\textwidth,file=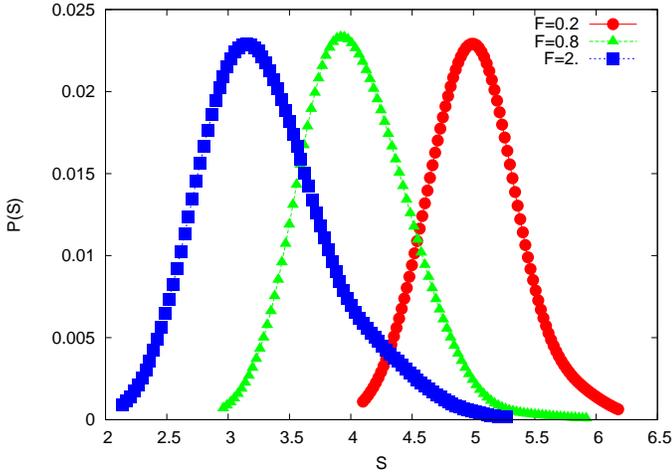}
\caption{  Probability distribution $P(S)$ of "entropy" $S$ defined by (\ref{entropia}), obtained by the
numerical simulations, for three different values of $F$, namely $F=0.2,0.8$ and $F=2.$. }
\label{fig4}
\end{figure}

Finally in Fig. (\ref{fig4}) we show the probability distribution $P(S)$, obtained by the
numerical simulations, for three different values of $F$, namely $F=0.2,0.8$ and $F=2.$. As one can see,
the maximum $P(S)$ is shifted toward small values of $S$ for increasing $F$, as we already know from
Fig.  (\ref{fig2}). Fig. (\ref{fig4}) shows that the fluctuations of $S$ about the mean are approximately
independent of $F$.

\bigskip

\begin{figure}
\centering
\epsfig{width=.50\textwidth,file=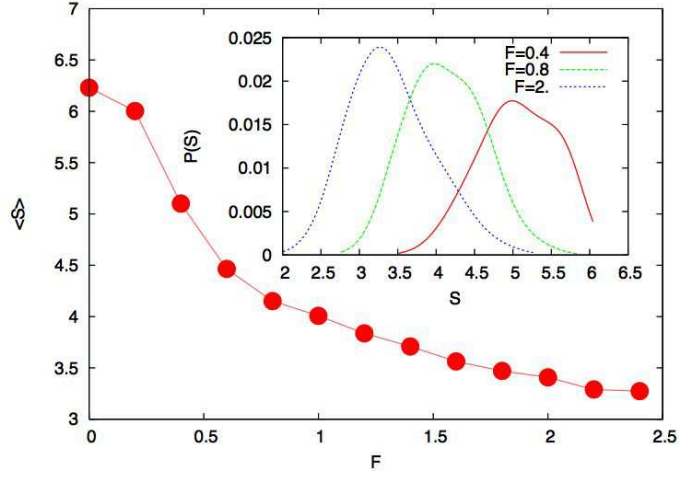}
\caption{ Computation of $\langle S \rangle$ as a function of $F$ for $\mu=1$. 
In the insert, we show the 
prrobability distribution $P(S)$, obtained by the
numerical simulations, for three different values of $F$, namely $F=0.4,0.8$ and $F=2.$}
\label{fig5}
\end{figure}

We now turn our attention to the case $\mu>0$ . We have performed numerical simulations for two growth rates,
 namely $\mu=1$ and $\mu=5$. We start by analyzing the results for $\mu=1$.
In Fig. (\ref{fig5}), we show the behaviour of $\langle S \rangle $ as a function of $F$, while in the
insert we show the probability distribution $P(S)$ for three values of $S$. Upon
comparing with Fig. (\ref{fig5}) against
Fig.s (\ref{fig2}) and (\ref{fig4}),  we see that a nonzero growth rate
$\mu=1$ does not change the qualitative behavior
of the system, in agreement with our theoretical discussions in the previous section.

It is interesting to look at the time averaged bacterial mass
  $\langle Z \rangle$ as a function of $F$. In Fig. (\ref{fig5}) we show $\langle Z \rangle$ and $\langle S \rangle/S_{max}$ as a function of $F$. 
For large $F$, when localization dominates the behavior of $c(x,t)$, $\langle Z \rangle$ is quite small, order $0.1$ of its maximum value, i.e. due to turbulence the population $c(x,t)$
only saturates locally at a few isolated points. The reduction in $\langle Z(t) \rangle$ tracks in $\langle S(t) \rangle$, but is much more pronounced.

\begin{figure}
\centering
\epsfig{width=.50\textwidth,file=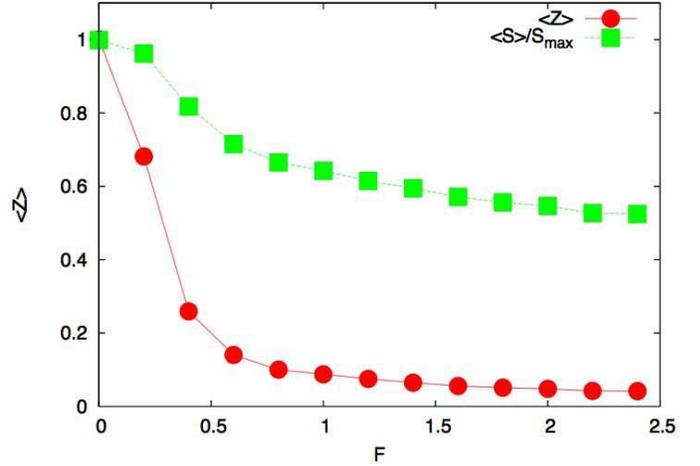}
\caption{ Computation of the total bacterial mass $\langle Z \rangle$ (normalized to $1$ at $F=0$)  for $\mu=1$ (red circles) and of $\langle S\rangle / S_{max}$ (green squares) as a function of $F$}
\label{fig5b}
\end{figure}

\begin{figure}
\centering
\epsfig{width=.50\textwidth,file=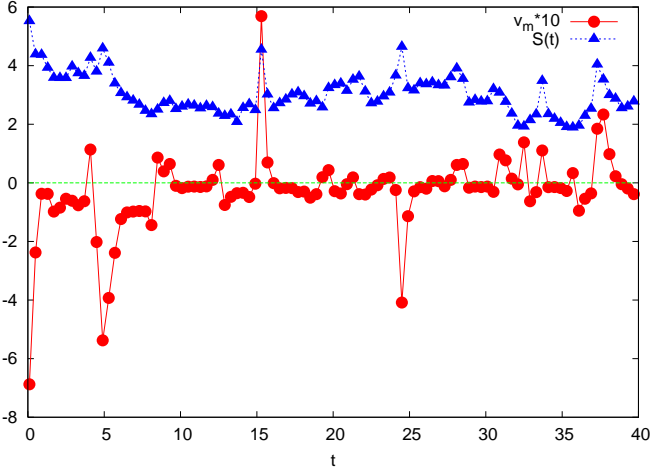}
\caption{ Same as in Fig. (\ref{fig3}) for $\mu=1$, non zero saturation term and $F=3.0$. }
\label{fig6}
\end{figure}

In Fig. (\ref{fig6}) we show $v_m(t)$ computed for the case $\mu=1$ and $F=3.$. As in  Fig. (\ref{fig3}),
we plot $v_m*10$ and $S(t)$. The qualitative behaviour is quite close to what already discussed for the case 
$\mu=b=0$. The whole picture for $\mu=1$, as obtained by inspection of Fig.s (\ref{fig5}) and (\ref{fig6}),
supports our previous conclusions 
that, as long as the systems is in a quasi-localized phase, the effect of $\mu$ 
in Eq.  (\ref{1d}) is almost irrelevant. Note that for the system to be
in the localized phase we must require that {\it both} conditions (\ref{11}) and (\ref{seconda}) must be 
satisfied. 

\begin{figure}
\centering
\epsfig{width=.50\textwidth,file=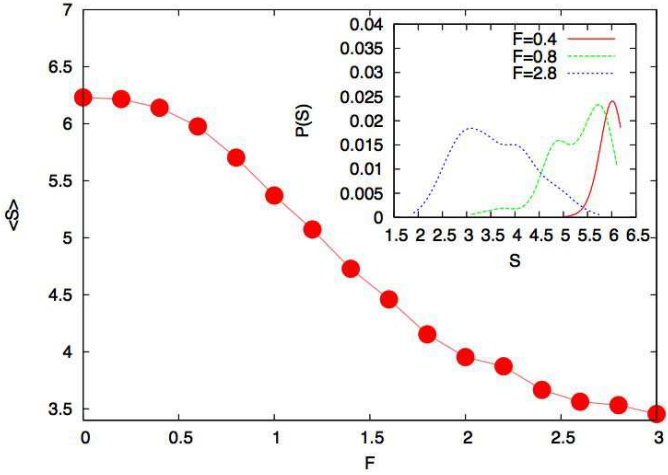}
\caption{ Same as in Fig. (\ref{fig5}) for $\mu=5$. }
\label{fig7}
\end{figure}

According to our interpretation, we expect that for increasing $\mu$  the whole picture does not change
provided $F$ is increased accordingly. 
More precisely, we expect that the relevant physical parameters are
dictated by the ratios in Eqs. (\ref{11}) and (\ref{seconda}). To show that this is indeed the case,
we show in Fig. (\ref{fig7})  the results corresponding to those in Fig. (\ref{fig5}) but now with
$\mu=5$ instead of $\mu=1$.

Two clear features appear in Fig. (\ref{fig7}). First the qualitative behavior of $\langle S \rangle$ with increasing
$F$ is
similar for $\mu=5$ and $\mu=1$. This similarity also applies to  the probability distribution $P(S)$ shown
in the insert of Fig. (\ref{fig7}). Second, there is a shift of the function $\langle S \rangle_{F}$ towards
large values of $F$, i.e. the localized/extended transition occurs for larger values of $F$ with respect to
the case $\mu=1$. This trend is in qualitative agreement with the condition (\ref{seconda}).

To make progress towards a quantitative understanding, we would like to use (\ref{prima}) and (\ref{seconda}) to
predict the shift in the localized/extended transition (or crossover) for increasing $\mu$. For this purpose, we need
a better indicator of this transition. So far, we used $S$ as a measure of localization: large values
of $S$ mean extended states while small values of $S$ imply a more sharply peaked probability distribution.
 For $\mu=b=0$, $S$ is the 
"entropy" related to the probability distribution $P(x,t)$, solution of eq. (\ref{1d}). Thus for $\mu=b=0$ we can
think of $S$ as the "entropy" and of the diffusion constant $D$ as the "temperature" of our system. This analogy suggests
we define a "specific heat" $C_s = D \partial S/\partial D$ of our system in terms of $S$ and $D$. 
After a simple computation we get using equations (\ref{7}) and (\ref{entropia_s}):
\be
\label{calore}
C_s(t) = \int_0^L dx P(x,t) [ log(P(x,t))  - \int_0^L dx P(x,t) log(P(x,t)) ]^2
\ee
After allowing a statistically stationary state to develop, we
then compute the time average $\langle C_s \rangle$ to characterize the "specific heat" of our system for
a specific value of $F$. It is now tempting to describe the localized/extended changeover associated with
(\ref{1d}) in terms
of the "thermodynamical" function $\langle C_s \rangle$. 
In other words, we would like to understand whether 
a change in the specific heat can be used to "measure" the extended/localized transition with increasing $F$. The above
analysis can be done also for $\mu>0$ , (when $c(x,t)$ is no longer conserved)
 by using $P(x,t) \equiv c(x,t)/Z$ where the "partition function"  $Z(t) = \int_0^L dx c(x,t)$.

\begin{figure}
\centering
\epsfig{width=.50\textwidth,file=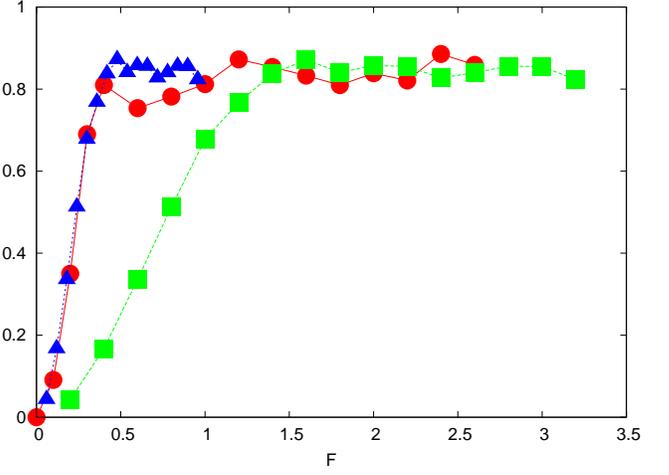}
\caption{ $\langle C_s \rangle$ as a function of $F$ for $\mu=1$ (red curve with circles) 
and $\mu=5$ (green thin curve). The blue line with
triangles is  $\langle C_s \rangle$ for $\mu=5$ plotted against $F/5^{3/4}$, for reasons discussed in the text. }
\label{fig8}
\end{figure}

In Fig. (\ref{fig8}) we show $\langle C_s \rangle$ as a function of $F$ for $\mu=1$ (red curve with circles) 
and $\mu=5$  (green thin curve). Two major features emerge form this figure. First,  
$\langle C_s \rangle$ is almost $0$ for small $F$ i.e. in the extended case. In the vicinity of 
 a critical value $F=F_c$, 
$\langle C_s \rangle$ shows a rapid rise to large positive values and it stays more or less constant upon increasing
$F$. The large value of $\langle C_s \rangle$ reflects enhanced fluctuations in $log(P(x,t))$ (analogous to energy fluctuations in equilibrium
statistical mechanics)
when the population is localized.
This behavior is in qualitative agreement with the notion of phase transition where (within mean field theory)
the specific heat rises after a transition to an
 "ordered state". Here, the "ordered state" corresponds to a quasilocalized, or sharply peaked probability distribution $P(x,t)$.
 Our numerics cannot, at present, distinguish between a rapid crossover and a sharp phase transition.

The second interesting feature emerging from Fig. (\ref{fig8}) is that $F_c$, the value of $F$ corresponding to the
most rapid arise of $\langle C_s \rangle_F$, 
 depends on $\mu$, as predicted by
our theoretical considerations. Indeed, as shown just below
Eq. (\ref{seconda}), we expect that $F_c \sim (\mu)^{3/4}$. 
To check  this prediction, we plot in Fig. (\ref{fig8}) a third line (the blue line with
triangles) which is just $\langle C_s \rangle$ for $\mu=5$ plotted against $F/5^{3/4}$. This rescaling is aimed
at matching the position of the extended/localized changeover
 for the same $F_c$ independent of $\mu$. The correspondence between the two curves
 in Fig. (\ref{fig8}) confirms our prediction.

\bigskip

Fig. (\ref{fig8}) shows that the statistical properties of $c(x,t)$ can be interpreted
in terms of thermodynamical quantities. How far
this analogy goes, is left to future research. The quantity
  $\langle C_s \rangle$ is in any case  a sensitive measure of the
extended/localized transition with increasing $F$.

\section{ Numerical simulations for $u_0 \ne 0$.}

As discussed in Sec. $2$, for fixed large $F$,
a mean background flow $u_0 \ne 0$ can eventually induce a transition
from localized to extended configurations of $c(x,t)$. More precisely, for large $F$, i.e. for 
$F$ large enough to satisfy (\ref{prima}) and (\ref{seconda}), the system will spend most of its time
in localized states provided the condition
(\ref{terza}) is satisfied. Thus, for large enough $u_0$ we expect a 
transition from quasi localized (i.e. sharply peaked)  to extended solutions. In this section we study this transition and check the 
delocalization condition in (\ref{terza}). 

\begin{figure}
\centering
\epsfig{width=.50\textwidth,file=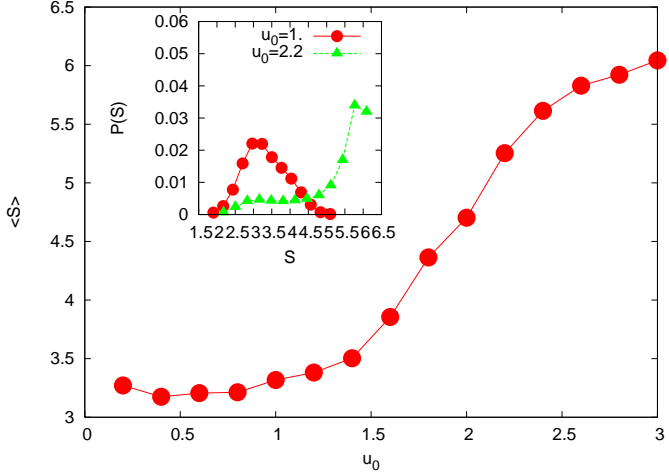}
\caption{ Plot of
$\langle S \rangle$ as a function of $u_0$  for $\mu=1$ and $F=2.4$.
In the insert we show the probability distribution $P(S)$ for
two particular values of $u_0$, namely $1.$ and $2.2$. }
\label{fig9}
\end{figure}

For this purpose we fix $\mu=1$ and $F=2.4$ which, according to our results in the previous section, 
correspond for $u_0=0$ to the case where localized states of $c$ dominate.
As before, we use $\langle S \rangle$ and of $P(S)$ to characterize 
the statistical properties of $c$ for different values of $u_0$. In Fig. (\ref{fig9}) we show
$\langle S \rangle$ as a function of $u_0$ while in the insert we show the probability distribution $P(S)$ for
two particular values of $u_0$. For $u_0 \sim 2$ we observe a quite strong increase of $\langle S \rangle$ , a signature of 
 a transition from predominately localized to predominately extended states. An interesting feature of $P(S)$ for
$u_0=2.2$ is the long tail towards small values of $S$. This means
that, occasionally, the system recovers  a localized concentration distribution, as if $u_0 = 0$. 

\begin{figure}
\centering
\epsfig{width=.50\textwidth,file=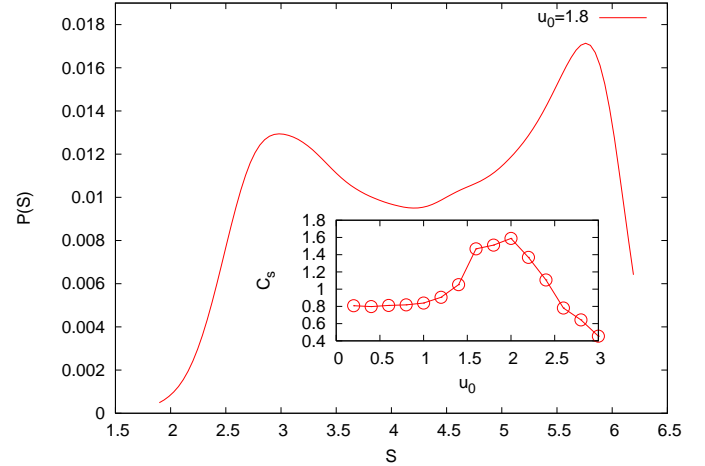}
\caption{ Plot of the probability distribution $P(S)$ for $u_0=1.8$, $\mu=1$ and $F=2.4$. In the
insert we plot $\langle C_s \rangle$ as a function of
$u_0$, where $C_s$ is computed using Eq. (\ref{calore}).}
\label{fig10}
\end{figure}

The most striking feature appears near the critical value of $u_0$ where the transition a sharp rise in $\langle S \rangle_{u_0}$ occurs. In Fig.
(\ref{fig10}) we show a two-peaked probability distribution
$P(S)$ for $u_0=1.8$, where the slope of $\langle S \rangle_{u_0}$ is the largest,  and in the insert we show $\langle C_s \rangle$ as a function of
$u_0$, where $C_s$ is computed using  Eq.(\ref{calore}). Let us first discuss the result shown in the insert of
Fig. (\ref{fig10}). The specific-heat like quantity  rises form $0.8$ at small $u_0$, shows a bump  where extended and localized states coexist, and
then drops to $0.4$ for $u_0$ large.
Note that the behavior of $\langle C_s \rangle$
is different from what we observe in Fig. (\ref{fig8}) suggesting a behavior reminiscent of a first order phase transition.
We  estimate $u_0 \sim 1.8$ as the critical value of $u_0$ where the behavior changes more rapidly.
At $u_0=1.8$ the
probability distribution is clearly bimodal, i.e. we can detect the two different phases of the system, one
characterized by highly localized states and the other characterized by extended states. Turbulent fluctuations
drive the system from one state to the other. The two maxima in $P(S)$ are suggestive of
two different statistical equilibria of the system. Note that $\langle C_s \rangle$ is once again
a good indicator of the transition from predominantly localized to predominantly extended states,
as discussed in the previous section.

\bigskip

For $u_0 \ne 0$, a straigthforward generalization of
Eq. (\ref{vm})  leads to the following results for the velocity of a maximum in $c(x,t)$, 
\be
\label{vmu0}
v_m = \mu Z \int_0^L dx (x_m-x) P(x,t)^2 + \int_0^L u(x,t) P(x,t) dx + u_0
\ee
One can wonder whether even for $u_0 > 0$, the localized regime of small $S$ shown in Fig. (\ref{fig10}) can
be still characterized by $v_m \sim 0$, thus representing a pinning of the concentration profile despite the drift  velocity $u_0$.
This question is relevant to understand whether the maxima for 
small $S$ in $P(S)$ shown in Fig. (\ref{fig10}) can be described using  ideas developed for quasi localized probability distributions
in Secs. $2$ and $3$. 
To answer the above question, we performed a numerical simulation with a time-dependent uniform drift
$u_0(t) = 1.6 + 0.8*cos(2\pi t/T)$ where $T = 10.0 $. Thus $u_0$ changes periodically
in time with an amplitude large enough to drive system from one regime to the other. If our ideas are reasonable, both
$S$ and $v_m$ will become periodic functions of time. 
In particular, as $S$l switches from small to large values, $v_m$ will go from $0$ in the localized regime to  a large positive
value in the extended phase. 

\begin{figure}
\centering
\epsfig{width=.50\textwidth,file=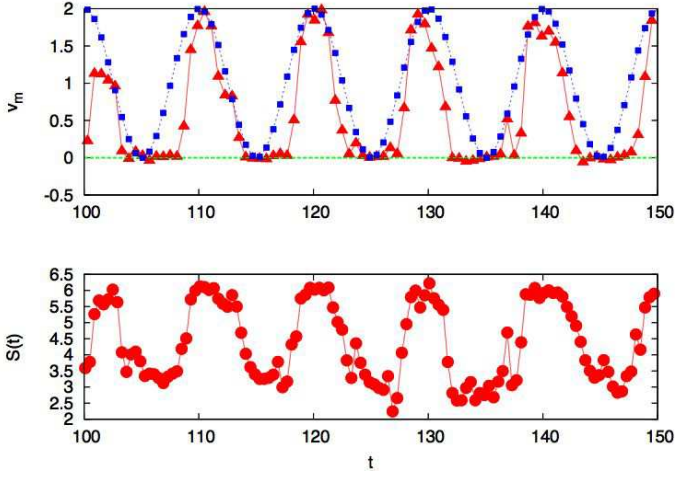}
\caption{ Time dependence of $S(t)$ (lower panel) and $v_m(t)$ (upper panel red triangles) for the case of a period mean
flow $u_0 = 1.6+0.8cos(2\pi t/T)$ with $T=10$, with the same conditions as in Fig (\ref{fig10}). The line with blue squares in the upper panel represents $cos(2\pi t/T)+1$. }
\label{fig11}
\end{figure}

\begin{figure}
\centering
\epsfig{width=.50\textwidth,file=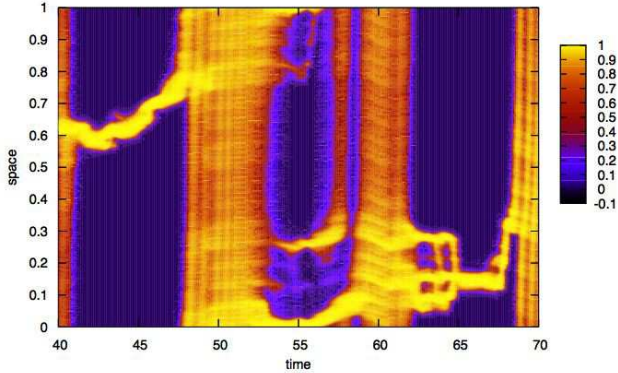}
\caption{ Contour
plot of $C(x,t)/Z$ (the horizontal axis is $t$ while the vertical axis is $x$) for the simulation
shown in Fig. (\ref{fig11}). }
\label{fig12}
\end{figure}

\begin{figure}
\centering
\epsfig{width=.50\textwidth,file=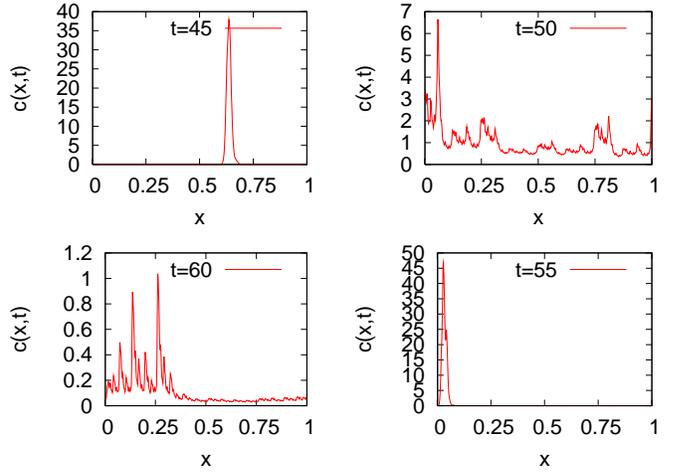}
\caption{ The figure shows four snapshots of $c(x,t)$ taken from FIg. (\ref{fig12})  at times $t=45,50,55,60$. At $t=45.$ and $t=55$, when $u_0=0.8$, the population
$c(x,t)$ is strongly localized while at $t=50$ and $t=60$, when $u_0=1.6$, $c(x,t)$ is extended. }
\label{fig12b}
\end{figure}

Fig. (\ref{fig11}) represents a numerical simulation  for both $S$ and $v_m$. In the upper panel
we plot $v_m$ (red line) and the periodic function $ cos(2 \pi t / T)+1$ (we add an offset of $1$ in order to
make the figure more readable). As one can see, $v_m$  indeed flattens out near periodically in time, and  increases to 
large positive values in synchrony with the external time-dependent drift velocity. 
In the lower panel, we plot $S$ as a function
of time; the graph  clearly shows a periodic switching between the two statistical equilibria.
A better understanding of the dynamics
can be obtained from Fig. (\ref{fig12}), where we show a contour
plot of the normalized bacterial concentration $c(x,t)/Z$ (the horizontal axis is $t$ while the vertical axis is $x$). Localized states can be observed in the vicinity
of $t=45,55$ and $t\sim 65$ i.e when $u_0$ is near its smallest value,  $u_0 \sim 0.8$. Localized states are stationary or at most slowly moving whenever
$u_0$ is small. During the period when $u_0$ is large, no localization effect can be observed. 
Fig. (\ref{fig12b})  we show four snapshots of $c(x,t)$ taken from FIg. (\ref{fig12})  at times $t=45,50,55,60$. At $t=45.$ and $t=55$, when $u_0=0.8$, the population
$c(x,t)$ is strongly localized while at $t=50$ and $t=60$, when $u_0=1.6$, $c(x,t)$ is extended.
The reason why $v_m \sim 0$ even for a small $u_0 >0 $ is quite simple: according to our analysis in Sec. $2$,
localized states will form near shifted zero velocity points with negative slopes even for $u_0 >0$. When $u_0$ is large enough, there is
no point where the whole velocity $u(x,t)+u_0$ is close to $0$. Every point in the fluid then moves in a particular direction, and the system develops
extended states.
Fig.s (\ref{fig11}), (\ref{fig12}) and (\ref{fig12b}) clearly support this interpretation.

\bigskip

\section{ Conclusions}

\begin{figure}
\centering
\epsfig{width=.50\textwidth,file=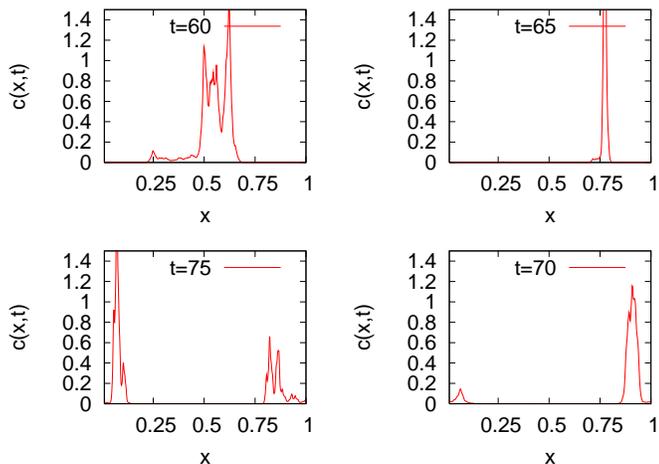}
\caption{ The figure illustrates one of the main results discussed in this paper: the spatial behavior of the population $c(x,t)$ subject to a turbulent velocity field.
The figure shows
four snapshots of $c(x,t)$ taken from a numerical simulation ($D=0.005$,$\mu=1$, $F=1.2$) at times $t=60,65,70,75$. The population $c(x,t)$ shows strongly peaked
concentration at time $t=65$ and $t=70$, while at times $t=65$ and $t=75$, $c(x,t)$ seems to be less peaked. The popultion $c(x,t)$ alternates strongly peaked solutions and more
extended ones. }
\label{fig12c}
\end{figure}

In this paper we have studied the statistical properties of
the solution of  Eq.(\ref{1d}) for a given one dimnesional turbulent flow $u(x,t)$. Fig. (\ref{fig12c})
 illustrates one of the main results discussed in this paper: the spatial behavior of the population $c(x,t)$ subjected to a turbulent field.
In particular, the figure shows
four snapshots of $c(x,t)$ taken from a numerical simulation ($D=0.005$, $\mu=1$,  $F=1.2$) at times $t=60,65,70,75$. The population $c(x,t)$ shows strongly peaked
concentration at time $t=65$ and $t=70$, while at times $t=65$ and $t=75$, $c(x,t)$ is more extended. The popultion $c(x,t)$ alternates strongly peaked solutions and more
extended ones
Our model is sufficiently simple
to allow systematic investigation 
without major computational effort. From a physical
point of view, the model can be interesting for compressible turbulent flows and whenever the field $c$ represents
particles (such as the cells of microorganisms) whose numbers grow and saturate while diffusing and advecting.
Our aim in this paper was to understand the statistical
properties of $c(x,t)$ as a function of the free parameters in the model. We developed in Sec. $2$ a simple theoretical
framework. Based on three dimensionless parameters, we have  identified three conditions which
must be satisfied for quasi localized solutions of (\ref{1d}) to develop, given by Eqs. (\ref{prima}),(\ref{seconda}) and (\ref{terza}).

\begin{figure}
\centering
\epsfig{width=.50\textwidth,file=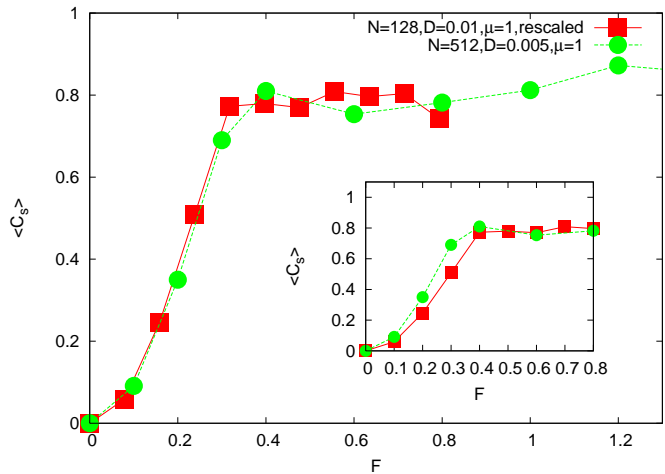}
\caption{  The "specific heat" $\langle C_s \rangle$ computed for $N=128$, $D=0.01$ and $\mu=1$ (red line with squares), rescaled according to
(\ref{gradient})  and (\ref{prima}),
 and compared
with the case $\mu=1$, $D=0.005$ and $N=512$ (green line with circles) discussed in the text. In the insert the same quantities are plotted without rescaling.}
\label{fig14}
\end{figure}

All  numerical simulations have been performed by using a grid resolution of $N=512$ points and a Reynolds
number $Re \sim 10^6$. Increasing the resolution will not change the numerical results provided the appropriate
rescaling on conditions (\ref{prima}),(\ref{seconda}) and (\ref{terza}) are performed, as shown in the following argument:
let us define $\delta x$ the grid spacing, i.e. $\delta x = L/N$, $\eta$ the Kolmogorov scale and $\epsilon$ the 
mean rate of energy dissipation, where $\eta = (\nu^3/\epsilon)^{1/4}$. The turbulent field $u(x,t)$ must be 
simulated numerically for scales smaller than the Kolmogorov scale. In the shell models, this implies that the largest
value of $k_n$ is much larger than $1/\eta$. The velocity gradient $\Gamma$ is of the order of $\sqrt{\epsilon/\nu}$.
If the grid spacing $\delta x$ is smaller than $\eta$, no rescaling is needed in using the theoretical considerations
derived in Sec. $2$, namely equations (\ref{prima}), (\ref{seconda}) and (\ref{terza}). On the other hand,
if $\delta x$ is larger than $\eta$, as in our simulations, the velocity gradient goes as 
\be
\label{gradient}
\Gamma \sim \frac{u(x+\delta x)-u(x)}{\delta x} \sim \epsilon^{1/3} (\delta x)^{-2/3}
\ee
Thus, by increasing the resolution, i.e. decreasing $\delta x$,  we increase the velocity gradients and the condition (\ref{prima}) may
not be satisfied unless we change $D$ or $F$ in an appropriate way. As an example of the above argument we show in
Fig. (\ref{fig14}) (insert) the value of $\langle C_s \rangle$ computed for $N=128$, $D=0.01$ and $\mu=1$ (red line with squares)
 and compared
with the case, used in the main text,  $D=0.005$,  $N=512$  and $\mu=1$ (green line with circles) already discussed in Sec. 3. For this particular
case, we can superimpose the two curves by multiplying $F$ for the $N=128$ case by a factor $0.8$ which
comes from equations (\ref{prima}) and (\ref{gradient}). The final result  agrees quite well with the
$N=512$ case as shown in the same figure.

Similar considerations apply for a non zero
mean flow $u_0$, where the localized/extended transition should occours for larger values of $u_0$ according to
(\ref{terza}). Finally, let us mention how we can predict the Reynolds number dependence of our analysis. According
to the Kolmogorov theory, a typical velocity gradient is
 $\Gamma \sim Re^{1/2}$. Therefore, the transition from extended to localized solutions
predicted by (\ref{prima}) can be observed provided 
 either $D\sim Re^{-1/2}$ or $F\sim Re^{1/2}$. Hence, by increasing the
Reynolds number,  the extended/localized transition eventually disappears unless the diffusion term $D$ is properly rescaled.
 
Following the theoretical framework discussed in Sec. $2$, we  introduced a simple way to characterized how well
$c(x,t)$ is localized  in space, namely using the entropy-like function (\ref{entropia}) to illuminate
the dynamics of the numerical solutions. The time average entropy $\langle S \rangle$ was used to characterize
the transition from extended to localized for increasing $F$ and from localized to extended solutions for increasing $u_0$. We also
found it useful to define a  "specific heat" $C_s$ by simply computing $D\partial S/\partial D$, where $D$ plays the
role of temperature in the system. Notice from Eq.  (\ref{prima}) that  the physics is controlled by an effective temperature $D/F$, i.e.
rescaling $F$ is equivalent to changing  $D$. 

The analogy between $F$ and some sort of effective temperature suggests that
the rapid rise in the time average $\langle C_s \rangle$, observed in Fig.
(\ref{fig8}) near a characteristic value  $F_c$, might indicate a  critical "temperature" or diffusion constant  $D_c$. Fig. (\ref{fig8}) highlights  the rapid changes in
$\langle C_s \rangle_{F}$  from extended to localized states in the system. It will be interesting to study 
 the behaviour observed in Fig. (\ref{fig8}) from a thermodynamic point of view.
As predicted by previous analytical studies \cite{nelson1} \cite{nelson2}, with time-independent velocity field,  a transition from localized to extended states has been observed 
by increasing $u_0$. The interesting feature is that near this transition, the system shows a clear bimodality in its 
dynamics, at least in the probability distribution $P(S)$, more indicative of a first order transition.

We are not able at this stage, to predict the shape of the probability distribution of $P(S)$ as a function of  external
parameters such as  $D$,$F$,$\mu$ and $u_0$. It would be valuable to understand better when a quenched approximation
(time-independent accumulation point in $u(x,t)$) 
is reasonably good for our system, especially in regimes where microorganism populations are nearly localized.
The reason why a quenched approximation may work is that
the localized regime is quasi-static, in the sense that
the solution $c(x,t)$ follows the slow dynamics of accumulation points 
where $u(x,t)=0$ with a large negative slope. A complete discussion of the validity of quenched approximation 
and analytic computations of $P(S)$ is
a matter for future research.

\begin{figure}
\centering
\epsfig{width=.50\textwidth,file=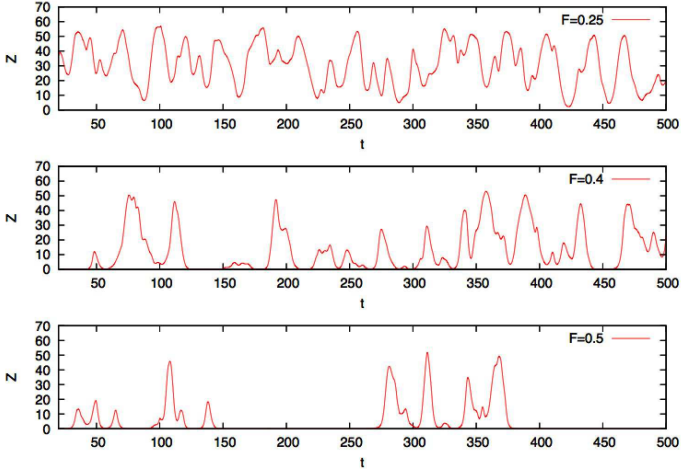}
\caption{ Numerical simulations performed with $\mu=1$ for $(7/16)L < x < (9/16) L$ and $\mu=-1/15.$ elsewhere for all $x$. We plot
the total number of microorganisms
$Z(t)$ for three different values of $F$, namely $F=0.25$ (upper panel), $F=0.4$ (middle panel) and $F=0.5$
(lower panel). }
\label{fig13}
\end{figure}

So far we have discussed the case of $\mu$ constant and positive. In some applications (both in biology and in physics) one
may be interested to discuss $\mu$ with some non trivial space dependence. An interesting case, generalizing the work in \cite{nelson1},
\cite{nelson2} and \cite{nelson3} is provided by the equation,
\begin{equation}
\label{oasis}
\partial_t c  +  \partial_x(Uc) = D \partial_x^2 c + \mu(x) c-bc^2, 
\end{equation}
with a turbulent convecting velocity field $U(x,t) = u_0 + u(x,t)$ and
where $\mu$ is positive on a small fraction of the whole domain and negative elsewhere.
In this case, referred to as the "oasis", one would like to determine when $\langle c(x,t) \rangle$
can be
significantly different from zero, i.e. when do the populations on an island or oasis survive when
buffeted by the turbulent flows engendered by , say, a major storm (see \cite{xx} for a treatment of space-independent random convection).
A qualitative prediction for $\langle c(x,t) \rangle$ results from the following argument: the extended and/or localization behaviour of $c$ 
depends on the ratio defined in Eq. (\ref{seconda}). For small $\Gamma$ (i.e. small $F$) the solution must be extended and therefore one can predict
that $c$ is significantly different from zero everywhere, 
wherever $\mu>0$. On the other hand, for large $\Gamma$, $c$ becomes localized. The
probability for $c$ to be localized in one point or another is uniform on the whole domain. Thus, if the region where $\mu>0$ is
significantly smaller then the region where $\mu<0$, $c$ should approach to zero for long enough time. 

In Fig. (\ref{fig13}),
we show a numerical simulation performed for a oasis centered on $x=L/2$, performed
with $\mu=1$ for $(7/16)L < x < (9/16) L$ and $\mu=-1/15.$ elsewhere for all $x$. Thus the spatial average of the growth rate is
 $L^{-1}\int_0^L dx \mu(x) = 1/15.$.
As a measure of $c$, we
plot its spatial integral $Z(t)$ as a function of time. The numerical simulations have been done with $D=0.005$, $N=512$ and $u_0=0$.
In Fig.(\ref{fig13}) we show three different values of $F$. Let us recall that, when $\mu=1$ everywhere, as is now the case for the oasis, 
for $F \ge 0.25$, the system exhibits 
a transition from extended states to localized states, as illustrated in Fig. (\ref{fig8}). As one can see,
for $F \ge 0.4$  the population  tends to crash as predicted by our simple arguments. It is however interesting to observe that the dynamics
of $c$ is not at all trivial. For $F=0.4$ and $F=0.5$, $c$ seems to almost  die and then recovers. Of course, our continuum equations neglect the
discreteness of the population. At very low populations densities, a reference volume can contain a fractional number of organisms and extintion events
are artificially supressed.
Fig. (\ref{fig13}) neverthelesst suggests an 
interesting feature of Eq. (\ref{oasis}) with space dependent $\mu$, worth investigating in the future.

\begin{figure}
\centering
\epsfig{width=.50\textwidth,file=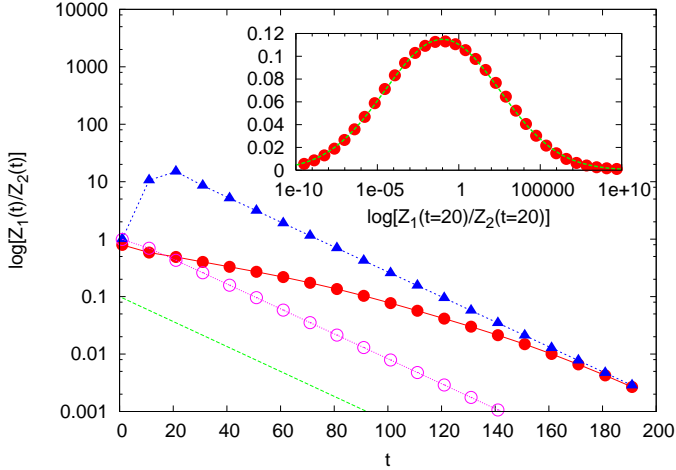}
\caption{ Numerical simulation of equations (\ref{c1}),(\ref{c2}) with $\mu_1=1.$ and $\mu_2=1.05$. We show the quantities $Z_1/Z_2$ (open circles)
for the case with no turbulence. Note that $Z_1/Z_2$ decays to zero as $exp(-(\mu_2-\mu_1)t)$ (green line) . When
turbulence is acting, the dynamics becomes more intermitent as shown by the behavior of $\langle Z_1\rangle/\langle Z_2\rangle$ (close red circles) and
and $\langle Z_1/Z_2 \rangle$ (solid triangles). The symbol $\langle ... \rangle$ means averaging over ensemble. In the insert, we show the probabiltiy distribution
of $log[Z_1(t)/Z_2(t)]$, at $t=20$, which is well fitted by a guassian behavior. }
\label{fig14b}
\end{figure}

Another interesting question tjhat deserves more detailed studies is the case two competing  species with densities $c_1(x,t)$ and $c_2(x,t)$. To illustrate the problem,
consider the coupled equations:
\begin{eqnarray}
\label{c1}
\partial_t c_1 + \partial_x (Uc_1) = D \partial_x^2 c_1 + \mu_1c_1(1-c_1)-\mu_2 c_1c_2 \\
\label{c2}
\partial_t c_2 + \partial_x (Uc_2) = D \partial_x^2 c_2 + \mu_2c_2(1-c_2)-\mu_1 c_1c_2
\end{eqnarray}
where $ \mu_2 > \mu_1$ and $0 < \delta \mu \equiv \mu_2-\mu_1 \ll \mu_1$. In this simplified model, Eqs. (\ref{c1}) and (\ref{c2}) 
 describe the dynamics of two populations in which a "mutant" density $c_2$ can out compete a wild type density $c_1$.
In particular, upon specializing to one dimension and  denoting $c_{sum}(x,t) \equiv c_1(x,t)+c_2(x,t)$, from Eq.s (\ref{c1},\ref{c2})
we obtain:
\begin{equation}
\label{c1c2}
\partial_t c_{sum} + \partial_x (U c_{sum}) = D \partial_x^2 c_{sum} + (1-c_{sum})(\mu_1c_1+\mu_2 c_2)
\end{equation}
Eq. (\ref{c1c2}) shows that, for $U=0$,  $c_{sum}=1$ is an invariant subset, i.e., if at $t=0$, $c_{sum}=1$, then $c_{sum}=1$ for any $t$.
The system has two stationary solutions, namely $c_1=1$,$c_2=0$ which is unstable, and $c_1=0$,$c_2=1$ which is stable. For $U=0$,
any initial conditions is attracted to the stable solution. It is easy to check that the asymptotic  time dependences in this subspace are of $Z_1 \sim exp(-\delta \mu t) $ and
$Z_2 \sim 1-exp(-\delta \mu t) $, where $Z_i \equiv \int dx c_i(x,t)$. 

In Fig.  (\ref{fig14b}) we show the result of two different numerical simulations of Eqs. (\ref{c1},\ref{c2}) with
$\mu_1 =1$ and $\mu_2=1.05$. The solutions have been obtained by using periodic boundary conditions, $L=1$, $D=0.005$ and a numerical resolution of $512$
grid points. The open circles represent the behavior of $log(Z_1(t)/Z_2(t))$  for $U=0$. As predicted by our simple analysis,
$Z_1/Z_2$ decays quite rapidly towards $0$ as $exp(-\delta \mu t)$ (dashed green line in Fig. (\ref{fig14b}) . Note that $\delta \mu = 0.05$, corresponding to a characterstic time
$1/\delta \mu \sim 20$.  

For $U \ne 0$, however,  the time behavior is quite different. In particular, we choose $u_0=0$ and allow convection by a strong turbulent field with $F=0.8$. In Fig. (\ref{fig14b}), 
the red circles
refer to $log(\langle Z_1(t) \rangle/\langle Z_2(t) \rangle)$ while the  blue triangles refer to $log(\langle Z_1(t)/(Z_2(t) \rangle$. The symbol $\langle ... \rangle$ is 
the ensemble average over $100$ realizations of the turbulent field, with the same initial conditions
\begin{eqnarray}
c_1(x,t=0) = 1 \ \  c_2(x,t=0)=0  \ \ for \ \ \ \ 0<x<\frac{L}{2} \\
c_1(x,t=0) = 0 \ \  c_2(x,t=0)=1  \ \ for \ \ \  \frac{L}{2} <x< L 
\end{eqnarray}
While the asymptotic states are still the same as for the case $F=0$ (the stability
of the stationary solutions does not change), the population $c_1$ decays on a time scale  longer than the $F=0$  one (i.e. $1/\delta \mu$).
The rather large difference  between 
$\langle Z_1(t) \rangle/\langle Z_2(t) \rangle$ and $\langle Z_1(t)/Z_2(t) \rangle$ is due to strong fluctuations in the ensemble. 
To highlight these fluctuations,
we show
in the insert of Fig. (\ref{fig14b}) 
the probability distribution $P(R)$ of the logarithmic ratio 
$ R(t)\equiv log(Z_1(t)/Z_2(t))$ computed at $t=20$, which is well fitted by a gaussian distribution with a rather large variance. This implies that
the ratio $Z_1/Z_2$ is a strongly intermittent quantity. To explain such a strong intremittency, note that the initial time behavior of the system strongly depends whether one of the two
populations is spatially extended while the other sharply peaked.
When the population $c_1(x,t)$ is extended while $c_2(x,t)$ is sharply peaked, the ratio $Z_1/Z_2$ becomes initially quite large. On the other hand,
when $c_2(x,t)$ is extended and $c_1(x,t)$ sharply peaked, $Z_1/Z_2$ is very small. For long enough times,  the two populations  become correlated in space (by clustering and competing
at the same accumulation points of $u(x,t)$)
 and the ratio $Z_1/Z_2$ eventually decays
according to the expected behavior $exp(-\delta \mu t)$. Note that the characteristic turbulent mixing times in our simulations are much longer that the characteristic doubling times
of the microorganisms, $\sim 1/ \mu_1$ and $\sim 1/\mu_2$. This is the opposite of the situation in many microbiology laboratories, where organisms in test tubes are routinely mixed 
at a rapid rate overnight
at Reynolds numbers of the order $10^3$. The situation studied here can, however, arise for microorganisms subject to turbulence in the ocean. 

We close with comments on generalization to more that one dimension. When
"turbulent velocity field" $\vec{u}(x,.t)$ can be represented as $\nabla \Psi(x,t)$, with a suitable $\Psi$, most of the results discussed in 
this paper should be valid. However, in a real turbulent flow in higher dimensions, whether compressible or 
incompressible, the velocity field is not irrotational. For a real turbulent flow, we believe the localization
discussed here will be reflected in a reduction of the space dimensions in the support of $c(\vec{x},t)$. For instance,
in two dimension, we expect that $c(\vec{x},t)$ will become large on a one-dimensional filament while in three dimension $c(\vec{x},t)$  localizes on a two dimensional
surface. For a review of related effects for biological organisms in oceanic flows at moderate Reynolds numbers see \cite{review}

Following the multifractal language, there may be a full spectrum of dimensions which may characterize the
statistical properties of localized states. It  remains to be seen  whether a sharp crossover (or an actual phase transition) similar to what has been shown in
Secs.  $3$ and $4$, will  be observed in more than one dimension.

\end{document}